\documentclass{article}

{}
{}
{}

\usepackage{amsmath} 
\usepackage{amsthm} 
\usepackage{cite} 
\usepackage{hyperref} 
\usepackage{graphicx}
\usepackage{algorithmic} 
\usepackage{hyperref}
\usepackage{authblk}
\usepackage[margin=1.2cm]{geometry}
\usepackage{multicol}
\usepackage{blindtext}
\usepackage{epstopdf}
\usepackage{color}
\usepackage{float} 

\usepackage{listings} 

\usepackage{todonotes}

\begin{document}

\title{\textbf{Trustchain -- Trustworthy Decentralised Public Key Infrastructure for Digital Credentials}}

\author[1]{Tim Hobson}
\author[2]{Lydia France}
\author[3]{Sam Greenbury}
\author[4]{Luke Hare}
\author[5]{Pamela Wochner}

\affil[ ]{
\vspace{-5mm}
\begin{center}
The Alan Turing Institute, London, UK
\end{center}
\vspace{-2mm}
thobson\textsuperscript{1}, lfrance\textsuperscript{2}, 
sgreenbury\textsuperscript{3},
lhare\textsuperscript{4}, 
pwochner\textsuperscript{5}\hspace{-0.2mm}@turing.ac.uk 
}
\date{}

\setcounter{Maxaffil}{0}
\renewcommand\Affilfont{\itshape\small}

\newcommand{\Sam}[1]{\textcolor{blue}{#1}}

\maketitle

\begin{abstract}
\noindent The 
sharing of 
public key information is central to the digital credential security model, but the existing Web PKI with its opaque Certification Authorities and synthetic attestations serves a very different purpose.
We propose a new approach to decentralised public key infrastructure, designed for digital identity, in which 
connections 
between legal entities 
that are represented digitally correspond to genuine, pre-existing 
relationships between recognisable institutions.
In this scenario, users can judge for themselves the level of trust they are willing to place in a given chain of attestations.
Our proposal includes a 
novel mechanism for establishing a root of trust 
in a decentralised setting 
via 
independently-verifiable timestamping. 
We also present a reference implementation
built on open networks, protocols and standards.
The system has minimal setup costs and is freely available for any community to adopt as a digital public good.
\end{abstract}

\begin{multicols}{2}

\section{Introduction}

Digital identity systems come in many guises, each design making a different set of trade-offs between diverse and competing factors including security, reliability, resilience, accessibility and scalability, each of which must be judged not only for its technological fitness but also its social and ethical implications. 
One axis for differentiation is the degree to which a central authority exercises control over the operation of the system and the data with which it operates. 

Centralised digital ID systems typically incorporate dedicated user registration processes, unique individual identifiers and the collection \& storage of personally identifiable information (PII) in a central database, often including biometric data \cite{mosip,banerjee2016aadhaar,anthes2015estonia}.
They are usually rolled out nationally and individual enrollment may be mandatory. 
These characteristics have raised concerns over potential unintended social exclusion and the risk of abuse by governing authorities (or adjacent actors), whether through disproportionate surveillance, individual profiling, or any number of undesirable consequences of ``function creep" \cite{mozilla2020bringing}.

Decentralised identity systems offer a means to address these concerns \cite{mole2023}. 
They typically rely on an existing infrastructure of open networks 
and protocols, assume voluntary participation and do not necessarily require any formal enrollment or entail any new data collection. 
Progress has been made in specifying open standards to support this approach, notably the closely-related W3C specifications for Decentralised Identifiers (DIDs) \cite{w3c2022did} and Verifiable Credentials (VCs) \cite{w3c2022vc}. 

While VCs can be used for individual identification, they are particularly suitable in cases where instead the goal is to prove set membership, by enabling certain attributes about an individual (e.g. their level of education or other qualification, employment status, country of residence, etc.) to be shared in a verifiable manner. 
Privacy can be further enhanced via ``selective disclosure'', a process by which a derivative Verifiable Presentation (VP) is used to disclose the minimum amount of information necessary to meet a given purpose.

The VC mechanism is predicated on the idea that verifiers are able to access the public keys of credential issuers in a trustworthy manner via their DIDs. 
Each DID is a unique identifier that can be ``resolved'' to obtain a DID document containing the subject's public keys together with associated metadata, which is assumed to be stored on a Verifiable Data Registry \cite{w3c2022did}. 
A specific mechanism for performing DID operations, including creation and resolution, is called a \textit{DID method}. 

DIDs therefore provide the building blocks for a decentralised public key infrastructure (DPKI) which is an essential ingredient in the VC operating model.
However, while it sets out the form and function of DIDs, the W3C standard leaves important 
questions unanswered 
and implementation details unspecified.
For instance, to what extent can decentralisation be reconciled with the inherently hierarchical trust model of credential issuers, holders and verifiers?
And what is to form the root of trust in such a hierarchical model?

In this paper we present a hybrid approach to DPKI, based on the DID standard and employing decentralised infrastructure that supports independently-verifiable timestamping. 
Our main contributions are: i) a mechanism for establishing a new public key infrastructure as a ``digital twin'' of real-world hierarchical trust relationships, making it suitable for digital identity applications, ii) a particular solution to the oracle problem which leverages verifiable timestamping to enable a user community to agree on one or more ``root'' DIDs, iii) a process for publishing ``downstream DIDs'' (dDIDs) to create chains of trustworthy decentralised identifiers leading back to the trusted root, iv) a mechanism for guaranteeing the completeness of DID revocation information, and v) a performant implementation of the proposed system as free and open-source software written in Rust.





\section{Related work}
By far the most widely used and well-known public key infrastructure is that used to secure communication on the World Wide Web.
Through the existence of centralised Certification Authorities, trust is conveyed via digital signatures attesting to the association of public keys with domain names, formalised in X.509 certificates~\cite{rfc4211}.
The Web PKI provides
trust in domain names for users accessing online services in spite of several well-known issues. 
A 2016 IETF memo highlights problems with certificate revocation status checking, 
issuance of ``surprising'' (i.e.\ unrequested) certificates and imperfect time synchronisation~\cite{draft-iab-web-pki-problems-02}.

While the Web PKI may remain an acceptable solution for domain name attestations, it falls short of an appropriate infrastructure for associating public keys with the physical identities of individuals and institutions. 
In part, this is due to the opacity of the Certification Authorities themselves, but the fundamental problem is that trust relationships represented by X.509 certificates are synthetic, in the sense that they are manufactured (on demand) for the specific purpose of 
sharing public keys. 
They do not attempt to capture genuine trust relationships that 
already exist 
between legal entities in the physical world.
These and other considerations have led to numerous attempts to build an alternative to the centralised Web PKI.

A
proposal for decentralised trust infrastucture introduced by Davie {\em et al.}\cite{davie2019trust} is the Trust-over-IP stack backed by the Trust-over-IP foundation.
The authors highlight the problems 
associated with 
centralised infrastructure, epitomised by the client-server paradigm, and discuss solutions derived from a parallel, layered decentralised technology stack 
with hierarchical governance.
While this work describes {\em what} the architecture of this decentralised infrastructure should look like, the question of {\em how} trust is to be conveyed between governance and technology stacks remains open.
Our work provides a compatible approach.

KERI, short for Key Event Receipt Infrastructure\cite{smith2019key}, is a system that aims to 
realise 
self-sovereign identity (SSI), a paradigm for  
online identity that allows users (individuals, organisations and
devices)
to manage their own identities without a centralised controlling authority.
KERI introduces a ledger-agnostic approach to decentralised identifiers and decentralised key management infrastructure that results in self-certifying identifiers (SCIDs).
An SCID cryptographically binds a public/private key pair to an identifier. 
A \textit{key event log} file is then used to keep a record of events such as key rotation, ensuring the verifiability of the correct key state.
Identifiers can be “delegated” to establish 
cryptographic links between
them and ultimately a hierarchy of identifiers can be created, 
but only under the assumption that the identifiers/keys of the root are known and trusted by all participants.

The ``Web of Trust''~\cite{zimmermann1995official} is a longstanding infrastructure that, in contrast to the Web PKI, aims to incorporate local trust assertions in a non-hierarchical way. 
Individual identity and public key associations are represented through digital signatures uploaded to keyservers where a path of digital signatures between entities needs to be formed. 
Given the ``small-world'' property of human interaction graphs, widespread trust may be attained as an emergent network property.
However, the Web of Trust has 
failed to achieve 
widespread adoption.
Additionally, a certificate spamming attack carried out in 2019 has led 
prominent 
authorities on cybersecurity 
such as Mozilla 
to abandon participation\footnote{\url{https://gpg.mozilla.org/}}.
Enthusiasm remains, however, and there are ongoing attempts to ``reboot'' the Web of Trust\footnote{\url{https://decentralized-id.com/workshops/rebooting-web-of-trust/}} model \cite{allen2015decentralized}.


In \cite{toorani2021decentralized} Toorani and Gehrmann recognise centralised certificate authorities as a 
cause of fragility in the traditional Web PKI, leading to ``compromises and operational failures''.
They suggest a blockchain-based PKI that aims to distribute trust among participating entities via a Web of Trust model. 
Public key registration, updates, and revocations are carried out through consensus among already-participating entities and revocations can be initiated by any member. 
However, the approach relies on the assumption that public keys can be validated during enrollment, but falls short of providing a mechanism for doing so.

A related contemporary approach to PKI, with a focus on interoperability and self-sovereign identity, is the European Blockchain Services Infrastructure (EBSI)\footnote{\url{https://ec.europa.eu/digital-building-blocks/wikis/display/EBSI/Home}}. 
This infrastructure aims to provide the institutions and citizens of EU member states with a decentralised ledger containing identifiers (with associated public keys), with a primary goal of facilitating cross-border interoperability. 
The ledger is maintained by a proof-of-authority mechanism, 
making it a {\em permissioned} blockchain, with write access determined by trusted authorities. 
A similarity with Trustchain is the reliance on socially-derived hierarchical trust relationships to support identity (the governance stack), while a key difference is the reliance on a centralised and permissioned ledger (the technology stack) that must be managed by the partner states.

\section{Design Principles}
The design of a trustworthy digital identity system must be based on 
sound principles that put the users and their rights at the centre of the system architecture. 
The following principles, based on those outlined by Goodell and Aste \cite{goodell2019decentralized}, are intended to inform and guide design choices. They also provide a framework for scrutiny of any proposed solution. 

\textbf{Privacy:}
Privacy is a human right \cite{assembly1948universal, europe1950humanrights} and privacy protections must be at the centre of any proposed system architecture. The individual must be in control of managing their identity and personal information in a multitude of contexts, including 
the ability to create multiple unrelated identities. 
Digital identity infrastructure must be designed in such a way that it does not facilitate the mass surveillance of individuals.

\textbf{Data minimisation:}
Only the necessary personal data for a specified purpose should be collected and retained. 
The design of digital identity systems 
should favour the sharing of verifiable attributes rather than 
full identities.
Where strong non-transferability of credentials is required it should be enforced interactively and at the edge of the network. 
Roles within the system should be isolated 
to prevent unnecessary data sharing (e.g. between credential issuer and service provider) which could be misused for profiling. 

\textbf{Trust:}
Trust in issuing authorities is inherent to any 
credential system, 
but the requirement for trust in the data infrastructure of the system should be minimised. 
Users at all levels should be able to 
understand where their trust is being placed.
Therefore synthetic trust relationships, established to facilitate technological ends rather than through genuine personal relationships and reputation, should be avoided. 

Institutional 
identity should be unique and public (while individual 
identities need not be either).
Consensual trust relationships between system participants should be realisable without requiring permission from an authority


\textbf{Transparency:}
Digital identity infrastructure
should be transparent and open to scrutiny. 
Code must be open source and should adopt open standards. 
Components on which the system depends must be realisable and identifiable; there should be no
hypothetical or vapourware dependencies.
Users must be aware whenever PII is shared or stored, and authorities should not be granted exceptional access (no ``backdoors'').

\textbf{Permissionless accessibility:}
A digital identity system should be inclusive, allowing all participants equal access without the need for permission from a central authority. 
Open protocols should be favoured over proprietary platforms, 
to avoid 
encouraging the formation of 
monopolies \cite{masnick2019protocols}. 
Participation should be voluntary and free from coercion. While opting out may result in reduced convenience, it should not incur any penalties.


    

\section{Trustchain System Design}
\label{sec:trustchain-system-design}

A verifiable credential with a digital signature enables its holder to to prove
certain individual attributes 
to which the credential issuer has attested.
%
It is presumed that the verifier not only trusts the issuing authority but is also able to 
reliably access their public key.
Imperfect knowledge of the
public key could lead the verifier to accept fraudulent credentials, since cryptographically valid signatures could be produced by a third party masquerading as the trusted issuer.

In a decentralised setting, the problem of how to share public keys in a trustworthy manner is a distinct challenge. 
This is the problem that Trustchain aims to address. 
%
In Sections \ref{subsec:downstream-dids} and \ref{subsec:root-did} we introduce two central ingredients upon which the design is based.
The first enables the construction of chains of trustworthy DIDs and the second provides the means to establish a root of trust, independently verifiable by all users of the system, at which DID chains terminate.
First, however, we describe the properties of the decentralised data layer used to store and share these identifiers.



\subsection{Verifiable Data Registry}
\label{subsec:verifiable-data-registry}

A central component in the W3C specification for Verifiable Credentials is the Verifiable Data Registry\footnote{\url{https://www.w3.org/TR/vc-data-model/\#roles}}, which is intended to store identifiers in the form of DID documents and document metadata. 
However the precise nature of this registry is left as an implementation choice. 
In particular there is no specification of exactly what should be verifiable, or of the verification mechanism.

Here we set out the specific properties required of the data registry in Trustchain (and in Section \ref{subsec:architecture} we exhibit a concrete example of such a registry). They are:
\begin{enumerate}
    \itemsep0em 
    \item Universal and permissionless read/write access.
    \item Independently-verifiable timestamping.
    \item Effective immutability.
    \item Scalability (with respect to DID operation throughput).
    \item Iterability.
\end{enumerate}

\noindent
The first property is necessary to meet our design objective of open accessibility. In particular we seek a system that is available for adoption by any user community and at any scale. 
The second plays an important role in the Trustchain security model, as explained fully in Section \ref{subsec:security}. 
Immutability guarantees that information written to the registry will be available unchanged in the future, without which properties 1 and 2 are not meaningful. 
Scalability is an obvious requirement, but is also a common weakness in decentralised systems. 
Finally, the iterability property ensures that we can traverse the entire registry, scanning for DID operations, after which we can be certain to have observed every operation that has ever been published. 
Coupled with verifiable timestamping, this property provides a guarantee to the user that they have the most up-to-date information available at any given time (including DID revocation notifications).

We emphasise that timestamps produced by the registry must be \textit{independently} verifiable.  
That is, the verification mechanism must be available to anybody and, to avoid circularity, must not require any special knowledge or trust in any third party, such as knowledge of a particular public key. 
%
Work on this problem originated with Haber and Stornetta \cite{haber-stornetta}, who proposed timestamping via a linked sequence of cryptographic hash digests, and was supercharged with the invention of Bitcoin \cite{nakamoto} which mixed in continuous proof of work (PoW), together with carefully aligned economic incentives, to achieve a fully decentralised timestamp server as a self-perpetuating distributed system.

The resulting mechanism has already been exploited in generic tools such as Open Timestamps\footnote{\url{https://opentimestamps.org/}}, and specifically for DIDs in certain DID method implementations\footnote{Notably the Identity Overlay Network (ION) on which our software depends; see Section \ref{subsec:architecture}.}.
Here we propose a novel use of verifiable timestamping to enable a user community to reliably share and agree on a particular DID that is to form the root of a hierarchical trust network.




\subsection{Downstream DIDs}
\label{subsec:downstream-dids}
A \textit{downstream DID} (dDID) is a W3C
Decentralised Identifier whose 
metadata includes a signature
from an entity attesting to both the identity of the
dDID subject and the validity of the information contained
in the corresponding DID document.
The attesting entity (or \textit{attestor}) is itself represented by an \textit{upstream DID} (uDID), 
through which its own public keys are made available.
This mechanism enables chains of trustworthy DIDs to be constructed, thereby allowing hierarchical networks of trust to be represented digitally on decentralised infrastructure.


\begin{figure}[H]
    \centering
    \includegraphics[height=6cm]{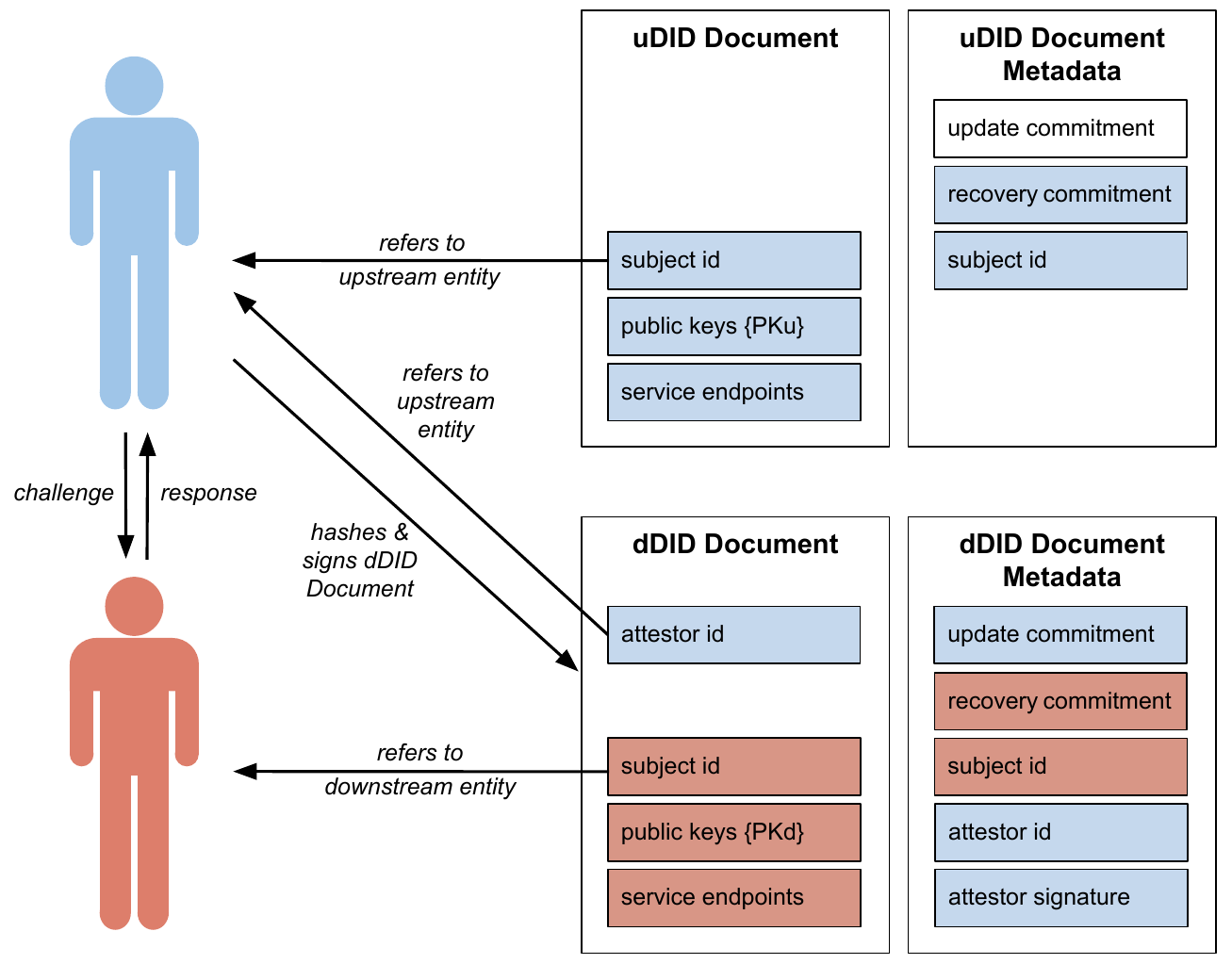}
    \caption{{\bf Upstream and Downstream DIDs}}
    \label{fig:ddid-schematic}
\end{figure}

\noindent Figure \ref{fig:ddid-schematic} illustrates the relationship expressed in a downstream DID and its essential information content.
To obtain a dDID, the subject first collects their own public key and service endpoint information into a candidate dDID document. 
(A \emph{service endpoint} is a network address, such as the URL of an HTTP API, that provides a means of communicating or interacting with the DID subject.)
The subject shares this candidate document with the upstream entity who then conducts a challenge-response procedure to verify that the subject is in possession of the private key counterparts to the given public keys. 
Since it is assumed that the two entities already share a well-established relationship, channels for this communication will certainly exist. 
But for convenience, and to ensure correctness, support for the challenge-response protocol is built into the Trustchain client software used by both entities.
Identity is also verified through the existing communication channels. 
%
%

When the upstream entity is satisfied with the quality of the information contained in the candidate dDID document, they compute its hash (digest), sign the hash and insert the signature into a DID document metadata file, which is then published along with the DID document itself.

The document metadata also includes an \textit{update commitment}. 
This is the hash of a secret value held by the attestor. 
Any attempt to update the DID document, if it is to be considered valid, must be accompanied by the pre-image of the hash. 
In this way, the attestor retains the right to revoke the dDID and can also perform renewals.
%
The dDID subject also holds a secret value from which a \textit{recovery commitment} is generated. 
This is included in the dDID document metadata, together with the update commitment, enabling the subject to regain full control of the DID identifier itself (without attestation) at any time. 

The result is a shared responsibility model, in which both
the subject and attestor are DID \textit{controllers}, meaning that both are authorised to make certain changes to the dDID document (the attestor has the right to revoke,
renew and update,
while the subject retains ultimate control of the identifier via the DID recovery mechanism).
Sharing the controller responsibility in this way ensures self-sovereignty for the dDID subject 
while providing the attestor with a unilateral revocation
and renewal
mechanism.

\subsection{Root DID}
\label{subsec:root-did}
The second ingredient is that of a \textit{root DID}, which represents the root of a trust hierarchy (analogous to a root certificate in the Web PKI).
The real-world entity (or entities) to which this root DID refers is assumed to be some authority that is recognisable to all users of the system.
For instance, if Trustchain were to be deployed at a national level, then a central government entity would be a natural choice for the root authority.

It is essential, of course, that all participants in the trust network (including credential issuers, holders and verifiers) are able to correctly identify the root DID, otherwise they could be tricked into accepting erroneous public keys and fraudulent credentials.
In a decentralised setting, where anybody is free to publish their own DIDs, how can users be confident that what they consider to be the root DID is indeed the genuine article?

This is a particular case of an issue known generally as the oracle problem, which crops up whenever a data registry is depended on as a single source of truth for information about the physical world \cite{caldarelli2020}.
While the integrity of the registry may be verifiable, the veracity of the information 
at the point of entry 
is not.
An additional source of trust is therefore needed if we are to have confidence in the data itself. 
And since our objective is to establish a root of trust, this additional assurance cannot take the form of a digital signature without undermining the entire scheme\footnote{It is sometimes said, correctly, that ``digital signatures do not create trust, they \textit{transport} it''. We would add that proof of work \textit{creates} trust.}.


In Trustchain, we propose a solution to this particular manifestation of the oracle problem by exploiting the data registry's capacity for verifiable timestamping. 
Here we assume this is achieved through continuous and cumulative proof of work (PoW), since no other known mechanism supports fully independent verification\footnote{The computational asymmetry between generating and verifying PoW also makes it effective against both Sybil and man-in-the-middle attacks.}. 
The idea is as follows.

Those members of the community who are to be represented in the root DID (being sufficiently prominent and trusted to play that role) produce a DID document containing a collection of their public keys, such that each member controls at least one of the counterpart private keys.
They then publish this DID document on the verifiable data registry, making it universally accessible and endowing it with a verifiable timestamp.
Having anchored the root DID in the proof-of-work chain, the next step is to widely publicise the fact by informing the whole community of its existence \textit{and} of the calendar date on which it was published.

The publicity should be as comprehensive as possible, so the special date itself becomes generally known and recognised. 
Any and all available channels of communication can be used for this purpose. 
Such channels will certainly exist, because the root entities are assumed to be prominent members of the community. 
And because a calendar date is a familiar and short piece of information, it can be shared over both electronic and non-electronic channels, including television \& radio broadcasts, newspapers, magazines, billboards, physical notice boards or even word of mouth.
The significance here is that the means for identifying the root DID can be shared widely via familiar and trusted channels, and without depending on any existing public key infrastructure.

Having established the root DID, the entities represented in it can begin issuing downstream DIDs to subordinate members of the community, who may themselves act as credential issuers and/or delegate to \textit{their} subordinates by publishing further downstream links in a DID chain.

Before the system is put into operation, a sufficient period of time must elapse to make the root DID effectively immutable. 
Any PoW chain can theoretically be retrospectively modified (or ``reorganised''), either by chance or by a deliberate attacker if they are able to out-perform the computational work done by the rest of the network. 
Fortunately, a material estimate of the cost of such an attack can be inferred from the network's aggregate hash rate, which itself can be accurately estimated from the regularly-published PoW digests. And because the work is cumulative, this cost increases linearly with the amount of time elapsed since the root DID was published.
A suitable waiting period can therefore be chosen to make the cost of attack prohibitive\footnote{Concrete estimates for the Bitcoin network are provided in Section \ref{subsec:practical-immutability}.}.

At the end of this waiting period, credentials can be issued by any of the entities represented in the network of DIDs emanating from the root. A relying party can then verify a credential by i) resolving the issuer's DID to obtain their public key, ii) tracing back through the chain of downstream DIDs, verifying the attestation signature at each step using the public key found in the next upstream DID, until they arrive at the root, iii) verifying the timestamp on the root DID, iv) verifying the signature on the credential using the issuer's public key, now known to be trustworthy.


\subsection{Security}
\label{subsec:security}
The Trustchain security model depends on the capacity of the data registry to support independently-verifiable timestamping.
In Section \ref{sec:implementation} we explain how this works in practice, but first let us consider why it is useful.

The primary security benefit is that it completely neutralises any attack\footnote{By an ``attack'', we mean any attempt to either modify the contents of the corresponding root DID document or to pass off a different DID document as if it were one referenced by the root DID.} on the root DID that takes place before or after the date on which the genuine root was published.
It is true that an attack could be carried out on the same date, by publishing alternative DIDs and/or DID documents which could later be confused with the root, but nevertheless, restricting the temporal attack surface to a particular 24-hour period is a major advantage.

In fact, coupled with the transparency of the data registry, the advantage is even greater. 
After the 24 hours have elapsed, all of the DIDs that were published on the same date are observable, and a simple check will reveal whether any ``fake'' root DIDs exist. 
If there are none, we can be certain that no attacker has published \textit{or will ever publish} a fake root DID that could be confused with our honest one.

To avoid potential DoS attacks in which fake root DIDs are published daily, on the off chance of coinciding with a genuine one, a short confirmation code (e.g. 3-6 characters) that uniquely identifies the root can be shared along with the publication date. 
In practice, the cost of the DoS attack then increases exponentially with the length of the confirmation code.


\vspace{-2mm}
\subsection{Features}

    

The design outlined above gives rise to a number of desirable features.
First let us consider those that are direct consequences of the decentralised nature of the data infrastructure. 

The system is designed to operate on top of existing peer-to-peer networks that are accessible to everybody, and we provide a free and open-source reference implementation (see Section \ref{sec:implementation}).
Therefore any user group can decide to deploy their own Trustchain instance without requesting permission and with minimal setup costs.
Each credential issuer is expected to run a full Trustchain node, but the system requirements are modest.
Users and verifiers can choose between running their own full node or instead opt for a ``light'' version of the software that runs on a mobile device and also functions as a credential wallet. 
(The security model for these light clients is explained below.)
No user registration is required. 
Operating costs are also low, but depend on the transaction costs associated with the immutable data registry. 
We give approximate figures in Section \ref{subsec:practical-immutability}.

Next we consider several features that are more specific to the design of Trustchain.

\vspace{6pt}
\textbf{Guaranteed-complete revocation data:}
Verifiable timestamping affords the ability to provide guarantees regarding the completeness of DID revocation information. 
It was observed earlier that certificate status checking is a significant issue in the Web PKI context. 
In Trustchain this problem is avoided because the data registry is the single, definitive (and fully iterable) source of DID information. 
No separate lookups are required to determine the revocation status of downstream DIDs. 
When taken in combination with the ability to verify DID timestamps, this has an important practical consequence: credential verifiers can establish, to a very high degree of certainty, that they have seen all available DIDs and all revocation information up to some recent (and \textit{precisely known}) time.
This feature could be used by verifiers to set internal policies or the terms of service level agreements.


\vspace{6pt}
\textbf{Constrained dDIDs:} 
Upstream entities in a DID chain, when issuing downstream DIDs, may wish to impose constraints that limit the permissions granted to their subordinates. 
This is analogous to reducing sub-CA privileges in the Web PKI, where DNS/URL name constraints are commonly used to tackle the problem of so-called ``surprising certificates'' \cite{draft-iab-web-pki-problems-02}.
In that context, however, the enforcement mechanism typically relies on best practice from browser vendors (or other monitors) to check for violations. 
The idea behind constrained dDIDs is to enforce these rules cryptographically.
Examples include:
\begin{itemize}
    \itemsep0em 
    \item limits on the total length of the dDID chain,
    \item name constraints on service endpoint URIs,
    \item constraints on allowed verification methods, or on other DID data/metadata parameters,
    \item constraints on the types of VCs that can be issued.
\end{itemize}
The first three examples relate to the subordinate's ability to further extend the DID chain. 
The last one relates to their ability to issue credentials, with the potential to significantly enhance the VC trust model by ensuring that issuers are only able to attest to attributes about which they are an appropriate authority. 
For instance, a bank may be allowed to issue credentials that attest to an individual's income threshold, or creditworthiness, but not a disability credential.

Possible approaches to implement constrained dDIDs include interactive schemes with multisignature verification methods and non-interactive schemes involving template certificates with sanitisable signatures. 
This is an area of active research.

\vspace{6pt}
\textbf{Rebasing:} 
Suppose two independent root transactions are created by two separate user communities. 
Each then builds a tree of trust relationships branching out from their respective roots
consisting of chains of downstream DIDs.

At some later date there may be a mutual agreement that one community should assimilate the other, by enabling chains of trust from (part or all of) tree $B$ to be traced back to the root of tree $A$.

This can be achieved by simply creating a dDID on tree $A$ that duplicates an existing DID in tree $B$ (which may be the root, but need not be). 
With this single act, all DIDs on tree $B$ that are downstream of the duplicated DID are automatically included in tree $A$, since there now exists a valid chain of trust to its root. Members of the community using tree $B$ would need to reconfigure their clients to accept the root DID timestamp of tree $A$, but no other changes are needed. Credentials already issued by entities in tree $B$ are now valid in tree $A$.

We refer to this operation as \textit{rebasing}, because it is analagous to the rebase operation found in the Git version control system. 
The act of rebasing updates the original DID on tree $B$, invalidating the original DID document content, because the most recent valid update overrides any earlier ones.
Note, however, that this does not invalidate any DID chains that were downstream of the re-based node on tree $B$, provided the same keys are included in the rebased DID document.

The option to rebase could ease adoption, as it enables different groups to begin using Trustchain independently without leading to permanently parallel and disconnected systems.

\vspace{6pt}
\textbf{Interoperability dDIDs:}
The concept of an \textit{Interoperability dDID} provides a means to support loosely-coupled federations across multiple, pre-existing digital ID systems in which VCs are issued either by an identity provider itself, or by a recognised (subordinate) issuer. This could enable credentials issued under one national ID system to be verified under another.

Before the two systems can become interoperable some sort of ``Trust Framework'' must be established between the Local and Remote entities, including a formal agreement to recognise credentials issued by the foreign system. Such frameworks necessarily take into account a variety of considerations (legal, regulatory, political, etc.) which are outside the scope of any technical solution. However, there remains the technical question of how a formal policy agreement is brought into effect in a live digital ID system. This is where the interoperability dDID can play a role.

A trust agreement between the two ID providers is embodied in a new downstream DID, signed by both entities in a joint attestation. This interoperability dDID represents a new combined entity, and the corresponding DID document includes the public keys already associated with each of the individual providers (in their own DID documents) for the purpose of VC issuance.

As soon as this new dDID is published, it is automatically downloaded and verified by users and verifiers in both systems. 
Now, when a holder presents their credential the verifier resolves the interoperability DID document to obtain the public keys with which they can verify its authenticity. The verifier also checks the chain of signatures on the interoperability dDID and finds a valid chain leading to the DID of their own ID provider, which is already recognised (and can be re-verified back to the trusted root if necessary).

Through this mechanism, credentials previously issued under either of the systems become instantly verifiable in the other one as soon as the interoperability dDID is published.

In order to achieve a similar arrangement, without the means of an open and shared public key infrastructure, bilateral or multilateral agreements between identity providers would need to be securely communicated to all users and relying parties, including updating lists of trusted public keys on their client devices. In that scenario device updates would become an attack vector and, in the absence of a universally accessible and verifiable ``ground truth'', the shared state across different participants in the system could easily become unsynchronised.

\section{Implementation}
\label{sec:implementation}

We present a reference implementation\footnote{\url{https://github.com/alan-turing-institute/trustchain}} of the design outlined in Section \ref{sec:trustchain-system-design}.
Written in Rust, it is intended not only as a proof of concept but as fully functional software capable of supporting a wide variety of practical use cases. 
This is possible because much of the technology stack already exists as operational open source software. 
In particular we make use of the \textit{Identity Overlay Network}\footnote{\url{https://identity.foundation/ion/}} (ION), a DID method maintained by the Decentralized Identity Foundation, and \textit{DIDKit}\footnote{\url{https://www.spruceid.dev/didkit/didkit}}, a cross-platform toolkit for VCs and DIDs from SpruceID.

The codebase is split into a core library, which is agnostic to the specific nature of the data layer, and an executable application that interacts directly with ION. 
The core library defines generic types and implements essential algorithms such as DID chain and timestamp verification.

The current software release supports root and downstream DID creation, resolution and verification together with VC issuance and verification. 
A mobile credential wallet, based on the Credible\footnote{\url{https://github.com/spruceid/credible}} app from SpruceID, is under development along with a Trustchain Server layer that responds to requests from mobile clients (as described in \ref{subsec:simplified-timestamp-verification}).

\subsection{Infrastructure}
\label{subsec:infrastructure}
In Section \ref{subsec:verifiable-data-registry} we set out the properties demanded of a verifiable data registry to support Trustchain as a higher-layer protocol.
Here we give a concrete example of just such a data registry, on which our reference implementation of Trustchain
is built. 

ION depends on two peer-to-peer systems for its decentralised infrastructure: the Bitcoin network and the distributed file system IPFS (or \emph{InterPlanetary File System}\cite{Benet2014}), which in combination deliver a data layer with the requisite features. 
The Bitcoin network provides the continuous and cumulative proof of work and permissionless write access (for publishing transactions) needed to support independently-verifiable timestamping. 
A Bitcoin transaction can contain up to 80 bytes of arbitrary data, more than enough to embed an IPFS content identifier (CID) which ION uses to reference the location of DID document data and metadata. 
Once stored on IPFS these files are immutable, since the CID is a cryptographic hash of the content itself. 

Scalability is also handled by the ION system, which inserts two layers of indirection between the embedded CID and the DID content, enabling several thousand DID operations to be anchored into the PoW chain in a single Bitcoin transaction.
In addition to enabling ample throughput, this implies minimal operating costs. With median Bitcoin transaction fees typically less than \$1 (USD) per transaction, batching of DID operations can be used to achieve a per-operation cost of a fraction of a penny.

The PoW itself is a (double) SHA-256 hash of the 80-byte header of a Bitcoin transaction block. Crucially, this header contains both a cryptographic commitment to all of the transactions in the block \textit{and} a timestamp field, comprising a 4-byte integer in Unix time format.


\subsection{Architecture}
\label{subsec:architecture}

\begin{figure}[H]
    \centering
    \includegraphics[height=5cm]{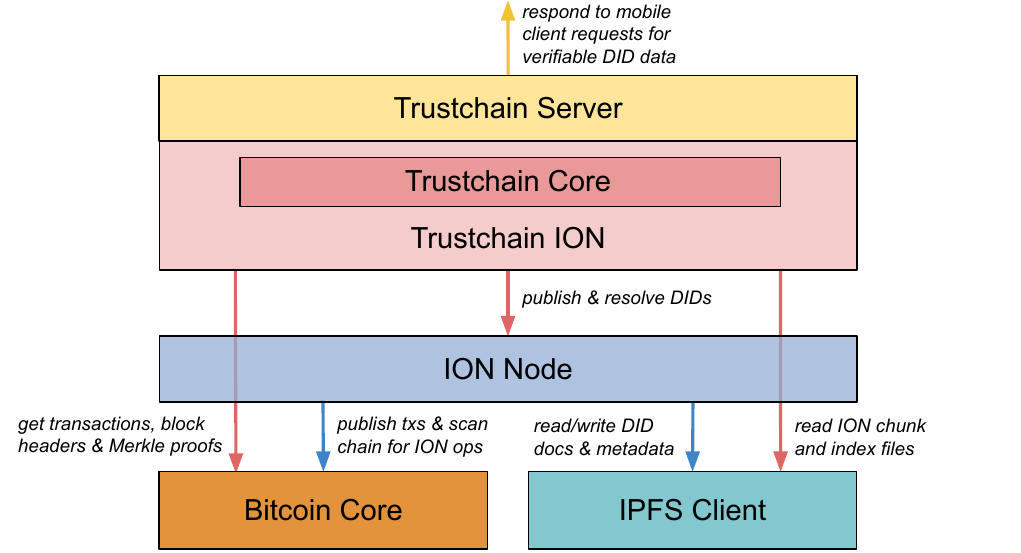}
    \vspace{-2mm}
    \caption{{\bf Trustchain functional architecture}}
    \label{fig:functional-architecture}
\end{figure}

Trustchain delegates DID operations to a locally-running node on the Identity Overlay Network (ION), which itself contains a client on each of the Bitcoin and IPFS peer-to-peer networks  (Figure \ref{fig:functional-architecture}). 
The former is used to publish Bitcoin transactions that anchor DID operations in the chain of PoW, and the latter reads and writes DID document data and metadata.

\noindent Trustchain also interacts directly with the Bitcoin and IPFS network clients in order to retrieve the data needed to independently verify timestamps associated with DID document content. 
This process is explained in Section \ref{subsec:practical-timestamp-verification}. These same bundles of verification data are also returned by the Trustchain Server layer in response to requests from mobile clients, enabling them to perform timestamp verification (see Section \ref{subsec:simplified-timestamp-verification}).
Therefore users of the Trustchain mobile app need not trust the server to supply accurate information because they can verify for themselves both the DID content and its timestamp.

\subsection{Practical immutability}
\label{subsec:practical-immutability}

Both of the peer-to-peer protocols on which ION depends contribute to the immutability of the Trustchain data layer.
In IPFS this is achieved through the mechanism of content-addressable storage, whereby files are retrieved via an address (the CID) which is a cryptographic commitment to the file content itself.
Bitcoin generates transaction IDs in a similar way, but consensus over which sequence of transaction blocks should be 
recognised depends on the ongoing proof of work performed by the network.
The protocol dictates that the chain exhibiting the greatest cumulative proof of work must be selected. Therefore, at any given time, it is impossible to guarantee that what is currently the ``strongest'' (i.e.\ most computationally expensive) chain will not be outpaced, and rejected, in the future.

Fortunately, the proof of work mechanism gives rise to a quantitative metric with which to judge the likelihood of a chain reorganisation. 
The relevant quantity is the cost of a deliberate attempt to modify the history of the chain (a so-called ``51\% attack'').

The first step is to estimate the network hash rate, that is, the rate at which computational work is being done by the network in aggregate. This can be done by observing the difficulty threshold imposed by the Bitcoin protocol and the rate at which valid blocks are being produced. Taken together with well-known performance characteristics of the specialised hardware used to perform these hashing operations, this implies a rate of energy consumption from which the cost-of-attack metric can be derived.

At the time of writing, the aggregate hash rate of the Bitcoin network is approximately 300EH/s and the theoretical cost (in USD) to modify the history of the chain, as a function of the temporal depth of the attack, is as follows\footnote{These figures are based on the total power estimate for the Bitcoin network and average electricity costs, provided by the Cambridge Bitcoin Electricity Consumption Index \url{https://ccaf.io/cbeci/index}. The estimates only consider energy consumption, ignoring the capital cost of acquiring mining hardware, and hence underestimate of the real cost of an attack.}:
\begin{center}
\begin{tabular}{ c|c|c|c|c } 
 \textbf{0} & \textbf{1 hour} & \textbf{1 day} & \textbf{1 week} & \textbf{1 month} \\ 
\hline
 $<\!\$1$ & \$700,000 & \$16.8m & \$117.6m & \$504m\\ 
\end{tabular}
\end{center}

\noindent It is very cheap to create honest timestamps but very expensive to create deceptive ones.


\subsection{Practical timestamp verification}
\label{subsec:practical-timestamp-verification}
While ION is a particularly convenient tool for performing DID operations, and its source code is open to audit, implementations of Trustchain need not, and should not, trust ION (or any other DID method) to perform its role faithfully. 
Instead, DID document timestamps can and should be verified directly. 
Most importantly, the timestamp on the root DID must be verified whenever a DID chain is assessed.

This process involves verifying a sequence of cryptographic commitments that tie the relevant DID document content (public keys and API service endpoints) to the timestamp inside a Bitcoin block header. 

\begin{figure}[H]
    \centering
    \includegraphics[height=5.5cm]{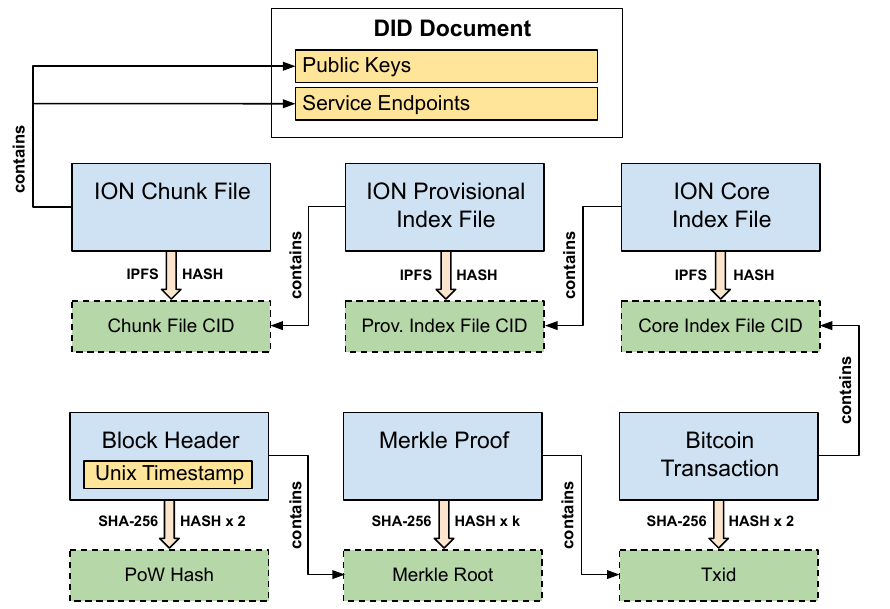}
    \caption{{\bf Timestamp verification} via a chain of cryptographic commitments (in green). Process inputs are in yellow.}
    \label{fig:timestamp-verification}
\end{figure}

\noindent Figure \ref{fig:timestamp-verification} illustrates the process, which takes two inputs: i) a DID document and ii) a 4-byte integer Unix time purporting to be the timestamp. It begins with the retrieval of three ION files stored on IPFS: the \textit{chunk file}, the \textit{provisional index file} and the \textit{core index file}.
The chunk file contains the DID content itself, so the first step is to check that all of the public keys and endpoints found in the DID document are present in that file. 
Then the chunk file is hashed with the IPFS hashing algorithm to obtain its content identifier (CID). 
As well as acting as its retrieval address on IPFS, the CID is also a cryptographic commitment to the file content, in the sense that it is infeasible to construct any content that differs from the original but that produces the same CID when hashed.

The next step is to verify that the chunk file CID is found inside the provisional index file. Then that file is hashed to obtain its CID. 
In turn, the core index file is checked to contain the CID of the provisional index file, before hashing to obtain \textit{its} CID. 
If all of these checks are successful then the CID of the core index file is also a commitment to the original DID document content.

When ION publishes a DID operation it records the core index file CID in a Bitcoin transaction, using a so-called \texttt{OP\_RETURN} script to embed arbitrary data. 
The verifier retrieves the transaction from the Bitcoin blockchain, confirms that it contains the core index file CID, and then hashes (twice) with SHA-256 to obtain the \textit{transaction ID} (Txid).

An efficient way to verify that this transaction is contained in a particular Bitcoin block is via a Merkle proof, which can be requested directly from the Bitcoin client wrapped inside the ION node. 
This Merkle proof consists of a branch of the Merkle tree of transactions, in which the leaf node is the Txid computed in the preceding step. 
By computing pairwise SHA-256 hashes on the set of commitments in the Merkle branch we eventually arrive at the Merkle root, which is a commitment to the transaction of interest, and therefore also to the original DID content found in the ION chunk file.

Finally, the Merkle root is contained in the Bitcoin block header which, when (double) SHA-256 hashed, produces the PoW hash digest for the block. 
Anyone running a Bitcoin network node can verify that this hash is contained in the cumulative sequence of PoW that distinguishes the Bitcoin blockchain. 
(In particular, it must satisfy the stringent difficulty requirement imposed by the Bitcoin protocol.) 
As well as verifying the presence of the Merkle root in the block header, a check for the candidate timestamp is made. 
If both are found, and the header is successfully hashed to produce a valid digest that can be found in the PoW chain, then the verifier has established \textit{with certainty} that a transaction committing to the DID document content was published in the block with the given timestamp. The immutability of the block itself was discussed in Section \ref{subsec:practical-immutability}



\subsection{Simplified Timestamp Verification}
\label{subsec:simplified-timestamp-verification}

Continuous and cumulative proof of work 
(which enables 
independently-verifiable 
timestamping) first appeared in Bitcoin, the peer-to-peer electronic cash system \cite{nakamoto}.
In the Bitcoin whitepaper the author presents a method called \textit{Simplified Payment Verification} (SPV) that enables lightweight clients (suitable for mobile devices) to efficiently confirm 
payments without needing to download the entire blockchain of transactions\footnote{In Bitcoin the use of SPV is discouraged for privacy reasons, but in the current context there is no such issue because transactions used to anchor DID operations do not transfer monetary value. Instead they form part of an open messaging protocol.}.
Instead, only the chain of 80-byte block headers is required, dramatically reducing bandwidth and storage requirements\footnote{At the time of writing, the entire Bitcoin blockchain is approximately 516GB in size, while the chain of block headers is only around 62MB.}.
The method takes advantage of the Merkle tree \cite{merkle} data structure used to arrange transactions inside a Bitcoin block. 
A so-called Merkle proof (which can be requested from a full Bitcoin node) is then sufficient to prove that a particular transaction is contained in a given block. The size of a Merkle proof increases at a rate of $O(\log(n))$ with respect to the number of transactions in the block.

Essentially the same method can be used to verify timestamps, with the important consequence that the Trustchain system can be accessed securely on a mobile device.
We refer to this process as \textit{Simplified Timestamp Verification} (STV).
The mobile Trustchain client omits the ION node (see Figure \ref{fig:functional-architecture}) and instead includes a lightweight Bitcoin client.
When verifying a DID timestamp it makes an HTTP request to a full Trustchain node\footnote{Trustchain Server exposes an API for such requests, and the service endpoint (URL) can be included in a DID document for easy discovery.} to obtain the bundle of verification data described in Section \ref{subsec:practical-timestamp-verification}.
It then performs the same timestamp verification procedure as a full node (see Figure \ref{fig:timestamp-verification}) to establish the validity of the root DID and therefore of any particular DID chain.

Because the chain of block headers is available locally, via the lightweight Bitcoin node, the mobile client does not need to trust the server to provide accurate data. 
A malicious (or faulty) server that responds with spurious data will be exposed by the verification procedure.

Note, however, that the mobile client has no way to check that the information provided by the server is complete. DID update or revocation information could be omitted without detection. To mitigate this risk the mobile client can request data from multiple servers.

\subsection{DID proof mechanics}
To provide a verifiable attestation mechanism between upstream and downstream entities, a cryptographic proof must be associated with each dDID document 
%
such that both are retrieved when the DID is resolved.
The proof itself relates not to the
dDID
subject, but to the dDID document, and therefore the appropriate location for this information is inside the dDID document \textit{metadata}.

Currently the W3C DID specification \cite{w3c2022did} does not formally provide a field for capturing proofs within document metadata (although the topic has received much discussion and we have raised this as a proposed extension\footnote{\url{https://github.com/w3c/did-spec-registries/issues/469}}). 
Our present implementation therefore contains the following workaround.

\begin{figure}[H]
    \centering
    \includegraphics[width=0.84\linewidth]{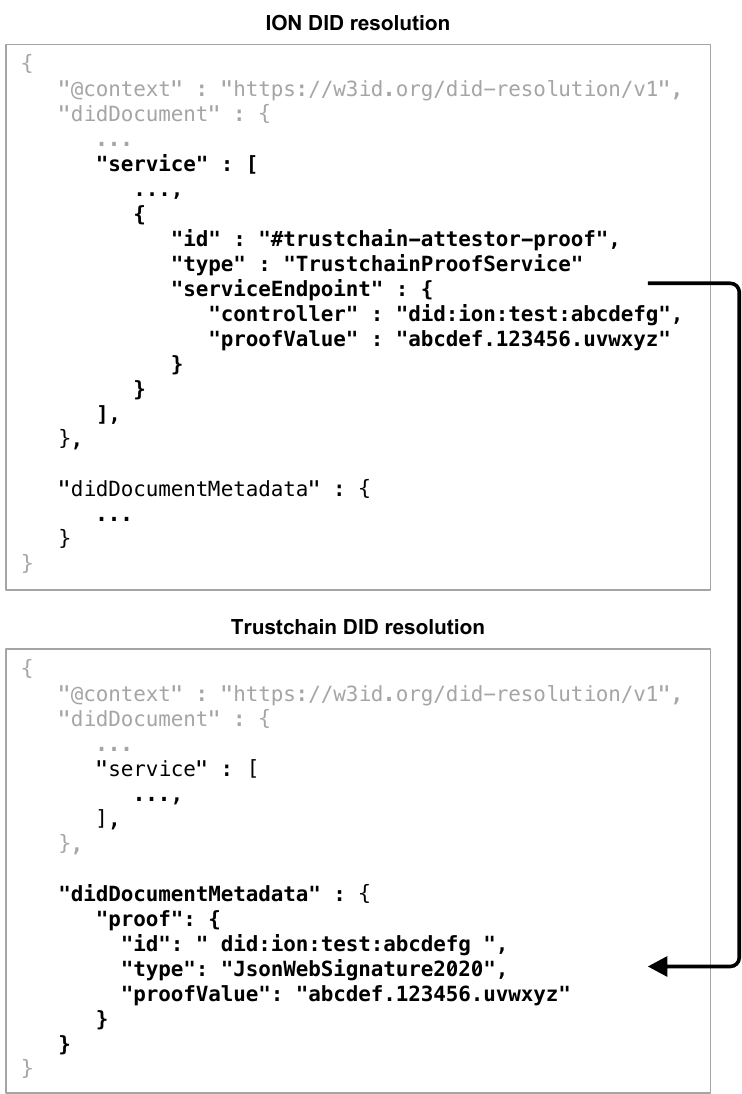}
    \vspace{-2mm}
    \caption{{\bf Trustchain proof mechanics.} An attestation proof is encoded as a service endpoint in an ION DID document (top) and transformed into DID document metadata (bottom) by the Trustchain resolver.}
    \label{fig:proof_mechanics}
\end{figure}

We introduce a
{\em trustchain-attestor-proof}
service ID that may be used in a DID document to denote 
a proof from an attesting entity. 
The proof itself may be accessible via a service (API) endpoint or directly inserted into the DID document.
When Trustchain resolves the DID, the contents of the proof service are moved into the document metadata and out of the DID document itself. 
An example of this transformation, from ION resolution to Trustchain resolution, is depicted in Fig.\ref{fig:proof_mechanics}.
\vspace{-3mm}



\section{Discussion}

Trustchain enables networks of trusted DIDs to be deployed at essentially arbitrary scale. 
It is for the community of users to decide the exact nature of the hierarchy, and therefore the types of institutions (or other legal entities) to be represented in the network.

If deployed at the national level, the central government would be represented in the root DID and would issue a downstream DID to each government department, for many of which there are obvious use cases. 
For instance, the Department for Health could publish a further dDID for each public hospital. 
This could be used to issue medical staff with credentials attesting to their level of qualification, enabling frictionless movement between hospitals.
Meanwhile the Department for Education may issue universities with dDIDs to enable verifiable digital degree certificates.
In both cases the chain of dDIDs would lead to the central government root, so a common knowledge of the root DID timestamp would be sufficient to guarantee security for these and many other use cases.

Trust networks may include entities from the public or private sector, or a mixture of the two.
For example, a financial regulator could issue a dDID whenever it grants a banking licence. 
Those licensed commercial banks could issue further dDIDs to the businesses to whom they provide financial services. 
Since the 
subject always retains control of the DID,
they cannot be locked-in.
And at each stage, the close relationship between superior and subordinate entities maintains trustworthiness in the chain.

A potential obstacle to adoption is the requirement that intermediate dDID holders must play a role akin to minor Certification Authorities, a responsibility they may be reluctant to accept. 
However these same organisations are likely to also be (or intend to be) credential issuers, which involves related tasks such as key management. 
They are also only required to issue dDIDs to entities with whom they already have an established relationship (otherwise they would not be the appropriate attesting authority).
The issue can be further mitigated by building automation for routine tasks, such as dDID renewal, into the Trustchain client software.

We have focussed on institutional identity, since the necessary ingredient to unlock the potential of VCs is the trustworthy identification of credential \textit{issuers}.
Holders generally need not be identified individually, and a variety of mechanisms are available to enforce non-transferability of credentials, such as one-time passwords and embedded photo ID, which can be selected according to context.
Nevertheless, downstream DIDs could equally be granted to individuals (e.g.\ as employees of a represented organisation), enabling further use cases.
This practice would not prevent those individuals from establishing other DIDs to represent alternative personas (if they so wished), and the absence of PII from DID document data circumvents privacy concerns.

Finally, a note of caution. Verifiable credentials, if widely adopted, 
could provide significant societal benefits
by dramatically reducing friction in the digital economy. 
However there remain important socio-political questions regarding the extent to which digital credentials should play a role in our daily lives.
Where they are used to improve efficiency, security and privacy in existing information sharing contexts, there is little risk of harm (provided a low-tech fallback option is retained to avoid exclusion).
But there are many other 
contexts,
particularly those that affect individuals' existing civil and political rights, 
in which the use of credentials (digital or otherwise) cannot be justified.
The existence of technological solutions must not be allowed to divert attention from this fact.

\section{Conclusion}

We have proposed a new, hybrid approach to public key infrastructure designed primarily to support verifiable digital credentials.
Our approach combines a model of hierarchical trust in issuing authorities and institutions with decentralised infrastructure.
This combination is justified by reference to a set of design principles including both practical and ethical considerations.
In particular, we aim to minimise trust in the underlying digital infrastructure, which should instead be \textit{verifiable}. 

The system is presented as both a theoretical design and a reference implementation, written in Rust and released as free and open-source software.
We also highlighted several promising features intended to improve security and promote adoption.
The setup and operational costs of the system are low, without compromising on security or resilience, because it is built upon existing, robust peer-to-peer networks.
It is also globally accessible and can operate at any social scale, making it freely available for any user community to adopt.


\section{Acknowledgements}
The authors gratefully acknowledge helpful discussions with Ed Chapman, Jon Crowcroft, James Geddes, Markus Hauru, Chris Hicks, Bryan Kumara, Carsten Maple, Vasilis Mavroudis, Ramesh Narayanan, Markus Sabadello and Pramod Varma.


This work was supported, in whole or in part, by the Bill \& Melinda Gates Foundation [INV-001309]. Under the grant conditions of the Foundation, a Creative Commons Attribution 4.0 Generic License has already been assigned to the Author Accepted Manuscript version that might arise from this submission.



\bibliographystyle{unsrt}
\bibliography{trustchain}
%







\end{multicols}

\end{document}